\documentclass[aps,prx,superscriptaddress,reprint,longbibliography]{revtex4-1}
\usepackage{graphicx}
\usepackage{xcolor}
\usepackage{siunitx}
\usepackage[normalem]{ulem}
\usepackage {amsmath}
\usepackage{bm}
\usepackage{color} 
\usepackage{comment} 
\usepackage[normalem]{ulem}
\renewcommand{\epsilon}{\varepsilon}

\begin{document}

\title{Network-forming phase separation of oppositely charged \\ polyelectrolytes forming coacervates in a solvent}

\author{Jiaxing Yuan}
\affiliation{Research Center for Advanced Science and Technology, University of Tokyo, 4-6-1 Komaba, Meguro-ku, Tokyo 153-8904, Japan}

\author{Hajime Tanaka}
\email{tanaka@iis.u-tokyo.ac.jp}
\affiliation{Research Center for Advanced Science and Technology, University of Tokyo, 4-6-1 Komaba, Meguro-ku, Tokyo 153-8904, Japan}
\affiliation{Department of Fundamental Engineering, Institute of Industrial Science, University of Tokyo, 4-6-1 Komaba, Meguro-ku, Tokyo 153-8505, Japan}

\date{\today}

\begin{abstract}
{The formation of coacervates through phase separation of oppositely charged polyelectrolytes (PEs) is critical for understanding biological condensates and developing responsive materials. Traditionally, coacervates are viewed as spherical droplets with growth dynamics resembling liquid-liquid phase separation. However, our fluid particle dynamics simulations incorporating hydrodynamic and electrostatic interactions challenge this perspective. Here, we find that oppositely charged PEs form a percolated network even in semi-dilute solutions, coarsening with a unique growth law, $\ell \propto t^{1/2}$. This self-similarity, absent for neutral polymers in poor solvents, arises because PEs in good solvents exhibit weaker, longer-range attractions due to spatial charge inhomogeneity under global charge neutrality. This results in a lower density of the PEs-rich phase and reduced interfacial tension. Increased charge asymmetry further slows network coarsening. Additionally, coacervate droplets initially display irregular shapes due to weak interfacial tension, transitioning slowly to spherical forms. Our research provides new insights into coacervate morphology and coarsening dynamics.}
\end{abstract}

\maketitle

\section*{Introduction}

Solutions containing oppositely charged polyelectrolytes (PEs) undergo phase separation when inter-PE attractions outweigh mixing entropy, leading to the formation of a coacervate phase enriched with PEs and a supernatant phase~\cite{muthukumar2023physics,van2011polyelectrolyte}. The formation of polyelectrolyte complex coacervates (PEC) opens avenues for creating responsive functional materials, particularly in drug delivery applications~\cite{ishihara2019polyelectrolyte,margossian2022coacervation}. PECs exhibit the capability to load and release drugs in response to environmental changes such as pH and ionic strength~\cite{ishihara2019polyelectrolyte,potas2020challenges}. Moreover, insights derived from investigations into phase separation of oppositely charged PEs offer understanding of the formation of membraneless organelle within biological cells~\cite{berry2018physical,tanaka2022Viscoelastic,wollny2022characterization}. 

 Recent advances in the study of coacervates have highlighted the importance of integrating experimental and theoretical methods to establish structure-property relationships~\cite{sing2017development,sing2020recent,rumyantsev2021polyelectrolyte,neitzel2021expanding}.
Considerable theoretical~\cite{radhakrishna2017molecular,sing2017development,sing2020recent,shakya2020role,chen2022complexation,chen2022driving,chen2023charge} and experimental~\cite{bohidar2005effects,spruijt2013linear,srivastava2016polyelectrolyte,li2020effect,sing2020recent,fraccia2020liquid,margossian2022coacervation} efforts have been devoted to understanding the phase behaviours of coacervation. Previous simulations have provided critical insights into the effects of salt concentration~\cite{radhakrishna2017molecular,chen2022driving}, polymer length and concentration~\cite{chen2022complexation,chen2023charge}, charge patterns~\cite{chang2017sequence,danielsen2019molecular}, solvent quality~\cite{li2020effect}, and chain stiffness~\cite{shakya2020role,yu2022isotropic} on coacervate structures. For example, Monte-Carlo (MC) simulations of coarse-grained models have replicated experimental observations of salt partitioning, demonstrating that excluded volume effects often cause salt to favour the supernatant phase~\cite{radhakrishna2017molecular}. Field-theoretic~\cite{danielsen2019molecular} and MC simulations~\cite{chang2017sequence} have proven effective in predicting coacervation with complex charge patterns. A recent simulation study by Chen and Wang resolved the debate on the driving forces of coacervation, revealing that under typical conditions for weakly to moderately charged PEs, solvent reorganisation entropy --- rather than the counterion release entropy --- is the primary driving force~\cite{chen2022driving}. A comprehensive review of coacervate phase behaviours is beyond the scope of this article due to the extensive body of literature. For a detailed overview, readers are referred to several excellent review articles~\cite{sing2017development,sing2020recent,rumyantsev2021polyelectrolyte,neitzel2021expanding}.

Despite these advancements, a noticeable gap still exists in the investigation of phase separation dynamics, which has so far received limited attention~\cite{chen2022driving,chen2023charge}. While the ultimate outcome of phase separation is dictated by equilibrium thermodynamics, gaining insight into the non-equilibrium dynamics is equally essential. From a theoretical standpoint, this presents a significant challenge, requiring the consideration of the complex dynamic interplay between long-range hydrodynamic interactions (HI) and electrostatic interactions~\cite{chen2022driving,chen2023charge,yuan2024hydrodynamic} in charged aqueous solutions. Nevertheless, understanding phase separation dynamics is particularly crucial since the slow relaxation dynamics may hinder the attainment of equilibrium in phase-separating soft matter systems~\cite{Tanaka2000Viscoelastic,tanaka2017phase}.

Generally, the evolution of phase separation patterns proceeds to reduce the free energy associated with the interfacial area, a process known as ``domain coarsening''~\cite{onuki2002phase}. 
Phase separation often proceeds self-similarly after forming the sharp domain interface, leading to the power-law domain coarsening. The characteristic domain size $\ell$ grows as $\ell \propto t^\nu$, where $t$ represents elapsed time and $\nu$ signifies the growth exponent. Classical exponents known for ordinary liquid-liquid phase separation (LLPS) include $\nu=1/3$ for droplet patterns, which are attributed to mechanisms such as evaporation-condensation or collision-coalescence (due to diffusive~\cite{siggia1979late} or Marangoni transport~\cite{shimizu2015novel}), and $\nu=1$ for bicontinuous (or network) patterns associated with the hydrodynamic pumping mechanism~\cite{onuki2002phase}.

Coacervates have been frequently observed as microdroplets in experiments~\cite{perry2015chirality,deshpande2019spatiotemporal,wollny2022characterization} where the domain coarsening proceeds through the collision-coalescence mechanism. For example, Chen and Wang investigated the coarsening dynamics of microdroplets during the phase separation of oppositely charged PEs, including hydrodynamic degrees of freedom, using dissipative particle dynamics (DPD) simulations~\cite{chen2023charge}. Their study assumed a dispersion of nanodroplets comprised of a polycation-polyanion pair as the initial state and revealed that under the condition of charge symmetry, the coarsening growth exponent remains at $\nu=1/3$, aligning with the classical value observed in ordinary LLPS~\cite{onuki2002phase}. However, as the degree of charge asymmetry increased, the coarsening dynamics exhibited deceleration~\cite{chen2023charge}.
Their initial condition of polycation-polyanion pairs is chosen to mimic droplet phase separation morphologies~\cite{perry2015chirality,deshpande2019spatiotemporal,wollny2022characterization}.

However, some experimental studies report the formation of porous networks during coacervation between oppositely charged PEs~\cite{vanerek2006coacervate,Wu2010,sadman2019versatile,murakawa2019polyelectrolyte,durmaz2020polyelectrolyte,baig2020sustainable,baig2020tuning,li2024design}. Generally, the non-equilibrium phase separation dynamics critically depend on the initial conditions or protocols used to induce phase separation. Rapid mechanical stirring, often used in experiments, tends to form aggregates in the initial state, which might impact subsequent coarsening. Conversely, experimental methods such as ion injection-retraction through membrane filters enable the initiation of phase separation in charged systems from a homogeneous mixed state in a non-perturbative manner~\cite{tanaka2005network,tsurusawa2017formation,tsurusawa2018direct,tateno2019numerical}, which mirrors potential biological phase separation processes. Indeed, a recent experimental study using an ion retraction method reported that network-forming phase separation of oppositely charged PEs leads to the formation of stable hydrogels~\cite{murakawa2019polyelectrolyte}. Consequently, a crucial question arises: what is the underlying physical mechanism behind the formation of network-shaped coacervates, rather than traditionally observed microdroplets, through phase separation of oppositely charged PEs?

In this work, we propose that the formation of network-like coacervates shares a common mechanism with ``viscoelastic phase separation" (VPS)~\cite{Tanaka2000Viscoelastic}. VPS was first discovered in neutral polymer solutions where the viscoelastic effects, arising from slower polymer motion relative to the fast deformation generated by phase separation, lead to the formation of a transient network~\cite{tanaka1993,Koyama2007}. This discovery challenges the conventional assumption that the minority phase should form droplets. This phenomenon has since been observed in various dynamically asymmetric mixtures undergoing demixing, in addition to polymer solutions~\cite{tanaka1993,Koyama2007}. These mixtures include colloidal suspensions~\cite{tanaka2005network,Bailey2007}, clay suspensions~\cite{shalkevich2007cluster}, protein solutions~\cite{banc2019phase}, and surfactant solutions~\cite{Iwashita2006self}.

Given that PEs should have slow dynamics, it is natural to expect that viscoelastic effects play a critical role in the phase separation of PEs, or coacervate formation. Nevertheless, previous coacervate studies have primarily neglected the viscoelastic effects, often regarding the coarsening of droplet coacervates as ordinary LLPS. Moreover, the presence of electrostatic interactions may introduce additional unique features into the VPS of PEs that are absent in neutral polymers~\cite{Tanaka2000Viscoelastic, yuan2023mechanical}. These considerations have motivated us to understand the viscoelastic effects in coacervate formation, based on numerical simulations properly incorporating two types of long-range many-body interactions: electrostatic interactions and hydrodynamic interactions.

Specifically, in this work, we address the following scientific questions: (i) Under what conditions does droplet-forming or network-forming phase separation occur? (ii) Is there self-similarity in the domain coarsening of network phase separation? (iii) What are the effects of electrostatic charges on network phase separation?

Our findings reveal that coacervate morphologies (networks versus droplets) highly depend on the initial state: Starting from a thoroughly mixed state, oppositely charged PEs spontaneously form a space-spanning network, even in semi-dilute solutions (volume fraction $\phi \gtrsim 1$\%). This implies that the experimentally observed droplet-shaped coacervates may result from imperfect mixing at the initial stage. Remarkably, the network coarsening follows an unconventional growth exponent of $\nu=1/2$ under charge symmetry conditions, which is fundamentally distinct from the classical value $\nu=1/3$ reported in droplet-forming phase separation of oppositely charged PEs~\cite{chen2023charge}. This $\nu=1/2$ behaviour resembles VPS of neutral polymers~\cite{yuan2023mechanical}, but it fundamentally differs as the self-similarity persists over an extended period without dynamic slowing down. This feature arises from the weak collapsing tendency of PE pairs in good solvents, unlike neutral polymers in poor solvents.

We also demonstrate that coacervate morphologies are significantly influenced by the PE volume fraction. We observe a transition from an interconnected network at a volume fraction of $\phi = 1.2$\% to isolated, irregularly shaped clusters at $\phi = 0.6$\%. This suggests that a volume fraction above approximately 1\% is required to initiate network formation. This crossover volume fraction can be further decreased with longer chains. Surprisingly, the resulting coacervate droplets in good solvents are irregularly shaped, indicating very low interfacial tension. This phase separation behaviour contrasts sharply with the classical spherical morphology observed in droplet-forming LLPS. We attribute this to the fact that phase separation occurring in good solvent conditions is driven by the weak spatial inhomogeneity of charges under the constraint of global charge neutrality.

As the charge asymmetry intensifies, the coarsening decelerates due to the net charge buildup on the network's surface, eventually leading to the dynamic slowing down of electrostatic origin. The network structure cannot be stabilised spontaneously in the VPS of neutral polymer solutions. Conversely, in the VPS of PE solutions, the network structure can be stabilised by introducing charge asymmetry. These findings indicate that adjusting charge asymmetry can effectively regulate network stability, with profound implications for creating stable, mesh-like biological condensates capable of withstanding mechanical forces and developing porous network materials for industrial use.

\section*{Results}

\subsection*{Charge symmetry condition}

\begin{figure}[t!]
  \centering
  \includegraphics[width=8.5cm]{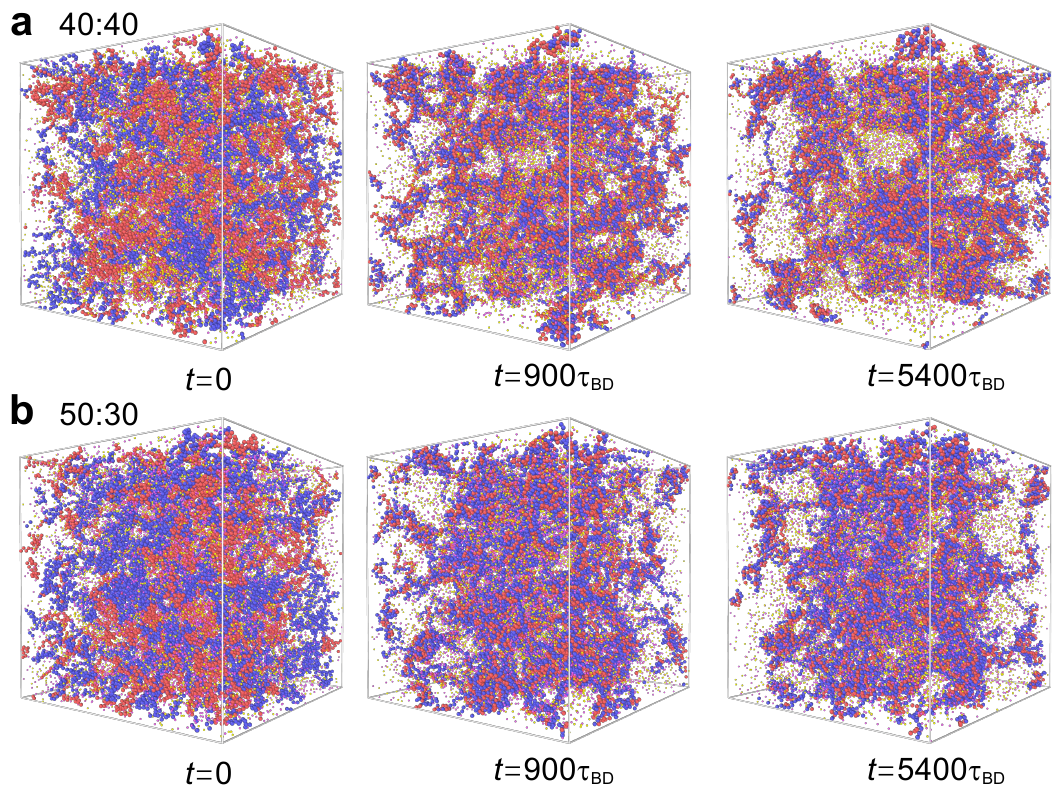}
  \caption{{Structural evolution during network-forming phase separation of oppositely charged PEs in a solvent.} Shown for a monomer volume fraction of $\phi\approx2.3$\% and a Bjerrum length of $l_{\mathrm{B}} = 3\sigma$. {\bf a,} Charge symmetric case ($N_\mathrm{a}=N_\mathrm{c}=40$). {\bf b,} Charge asymmetric case ($N_\mathrm{a}=30, N_\mathrm{c}=50$). Blue and red beads represent monomers of polycation and polyanion monomers. Counterions are depicted as smaller particles, with magenta and yellow beads indicating counterions associated with polycations and polyanions, respectively, for clarity. See Supplemental Movies S1 and S2 for corresponding visualisations of {\bf a} and {\bf b}.} 
  \label{fig:Fig1}
\end{figure}

We first examine the case of charge symmetry ($N_\mathrm{c}=N_\mathrm{a}=40$). Figure~\ref{fig:Fig1}a illustrates the temporal evolution of phase separation morphology following the complete mixing of polycations and polyanions. Remarkably, we observe that oppositely charged PEs spontaneously assemble into a space-spanning network, maintaining connectivity within our simulation timescales. This contrasts the formation of isolated droplet clusters observed previously from an initial state of nanodroplets containing a polycation-polyanion pair~\cite{chen2023charge}. This highlights the significant influence of initial conditions on morphology and suggests that the slow relaxation of interconnected PE chains kinetically drives the formation of a transient network, preventing its rapid transformation into droplet patterns.

The strand thickness and length of the network phase coarsen over time $t$. To characterise the coarsening behaviour, we monitor the temporal evolution of the characteristic wavenumber $\langle q \rangle$, defined as the first moment of the structure factor $S(q,t)$ (see Methods section for definition).
Interestingly, under the charge symmetry condition of $N_\mathrm{c}=N_\mathrm{a}=40$, we observe an unconventional power-law coarsening $\langle q \rangle\sim t^{-1/2}$ (Fig.~\ref{fig:Fig2}a, \ref{fig:Fig2}c, and \ref{fig:Fig2}e). This behaviour starkly contrasts the $\langle q \rangle\sim t^{-1/3}$ observed for droplet-like phase separation~\cite{chen2023charge}.
The power-law coarsening of $\langle q \rangle\sim t^{-1/2}$ remains consistent across different settings of the Bjerrum length $l_{\mathrm{B}} = 1\sim3\sigma$ (Fig.~\ref{fig:Fig2}a, \ref{fig:Fig2}c, and \ref{fig:Fig2}e), a behaviour notably absent in the limited prior studies on the dynamics of coacervate formation~\cite{liu2016early,liu2017structure,chen2022driving,chen2023charge}.

To assess the influence of HI on the coarsening law, we also conduct free-draining Brownian dynamics (BD) simulations (Fig.~\ref{fig:Fig2}b, \ref{fig:Fig2}d, and \ref{fig:Fig2}f; see Methods section for details of the BD method) using the same setup. Interestingly, we observe $\langle q \rangle \sim t^{-1/3}$ in BD simulations, despite qualitatively reproducing the network phase formation. This observation suggests a fundamental limitation in BD simulations, highlighting the essential role of HI in the non-equilibrium phase-separation dynamics.

We emphasise that the network-forming coarsening power laws of $\langle q \rangle\sim t^{-1/2}$ (with HI) and $\langle q \rangle\sim t^{-1/3}$ (without HI) are consistently observed in the VPS of neutral polymers~\cite{yuan2023mechanical}. This seems to suggest that coarsening dynamics under charge symmetry conditions share similarities with neutral polymer solutions, where the mechanical relaxation of the dense network phase controls the domain coarsening. This relaxation is limited by the solvent permeation flow through the narrow gaps between closely packed particles~\cite{tateno2021,yuan2023mechanical}. 

To affirm this scenario, we compare the structural relaxation time, $\tau_\alpha$, of the dense network phase with the characteristic domain deformation time, $\tau_\mathrm{d}$. Our findings show that $\tau_\alpha \gg \tau_\mathrm{d}$ in the cases we investigated (Fig.~\ref{fig:Fig2}g, \ref{fig:Fig2}h, and \ref{fig:Fig2}i; see Methods section for details on the estimation of $\tau_\alpha$ and $\tau_\mathrm{d}$). This result indicates that the slow particle rearrangement within the dense phase cannot keep up with the rapid domain deformation (including both shear and bulk deformation), leading to the activation of viscoelastic effects that leads to the formation of a transient network~\cite{tanaka1993,Koyama2007}. Thus, the coarsening dynamics are indeed controlled by the slow mechanical relaxation of the dense network phase~\cite{tateno2021,yuan2023mechanical}.

In this work, we focus on PE solutions in good solvent conditions. We anticipate that incorporating short-range hydrophobic attraction between monomers can further increase $\tau_\alpha$, making the experimental observation of network-forming VPS more likely. Conversely, reducing the electrostatic attraction between monomers via $l_\mathrm{B}/\sigma$ can decrease $\tau_\alpha$. Once $\tau_\alpha \ll \tau_\mathrm{d}$, the network-forming VPS~\cite{tanaka1993,Koyama2007} should be replaced by normal LLPS, where the coarsening dynamics are governed by the diffusive transport of materials~\cite{onuki2002phase}.

We note that the observed network structures should be regarded as a transient state. Whether the network structure (Fig.~\ref{fig:Fig1}a) transitions into coacervate droplets coexisting with a supernatant phase or forms stable networks in the long-time limit remains unclear. Directly observing the long-time morphological transformation presents inherent challenges that currently exceed our computational capabilities. Future experimental and numerical studies could provide critical insights into this unresolved question.

\subsection*{Charge asymmetry condition}

\begin{figure}[t!]
  \centering
  \includegraphics[width=8.5cm]{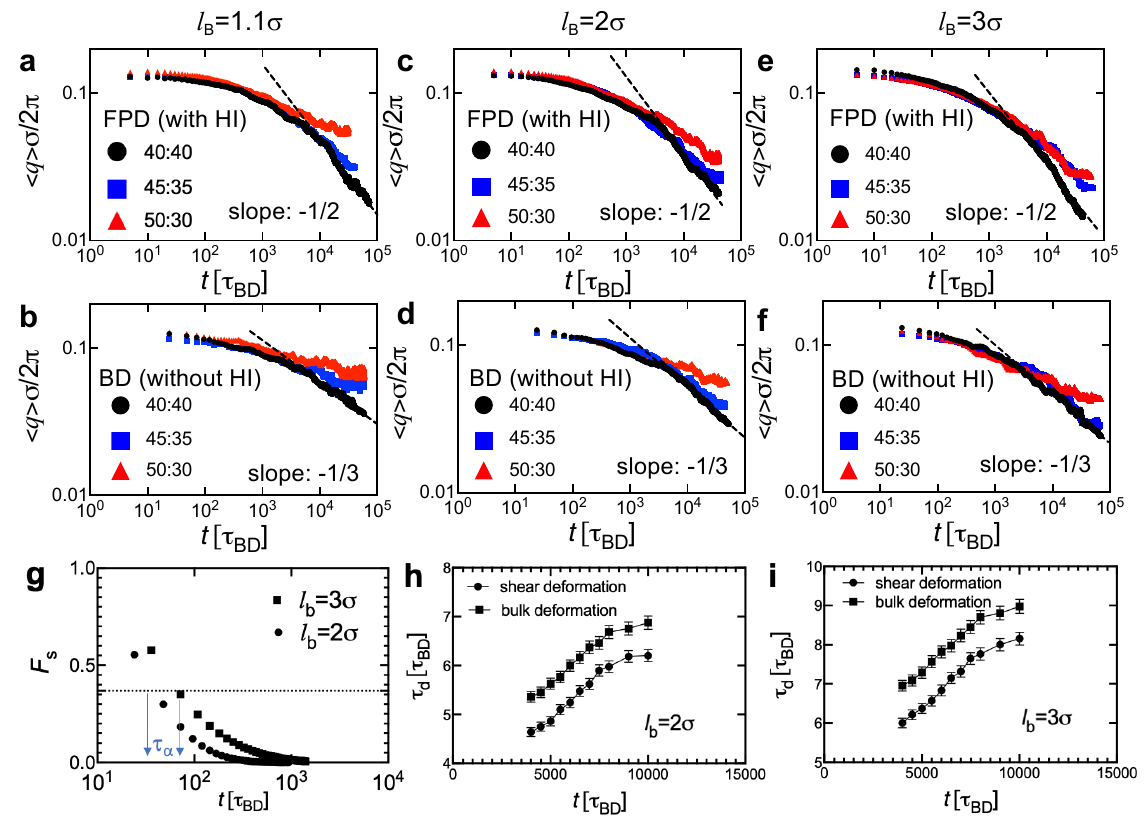}
  \caption{\footnotesize {Domain coarsening in semi-dilute solutions of oppositely charged PEs and characterisation of the timescales in phase separation.} {\bf a, c, e,} Temporal evolution of the characteristic wavenumber $\langle q \rangle$, defined as the first moment of the structure factor $S(q,t)$, for Bjerrum lengths $l_{\mathrm{B}} = 1.1\sigma$ ({\bf a}), $l_{\mathrm{B}} = 2\sigma$ ({\bf c}), and $l_{\mathrm{B}} = 3\sigma$ ({\bf e}) in FPD simulations. {\bf b, d, f,} Results obtained from free-draining Brownian dynamics (BD) simulations for the corresponding Bjerrum lengths. Error bars represent the standard errors calculated from four independent simulations (smaller than the symbol size if not visible). For charge symmetric conditions ($N_\mathrm{a}=N_\mathrm{c}=40$), domain coarsening follows $\langle q \rangle \sim t^{-1/2}$ and $\langle q \rangle \sim t^{-1/3}$ in FPD and BD simulations, respectively.  {\bf g,} Self-intermediate scattering function~$F_{s}({q}, t)$ for the dense phase of a binary charged PE solution with chain lengths $(N_\mathrm{c}, N_\mathrm{a})=(40,40)$ at Bjerrum lengths $l_{\mathrm{B}} = 2\sigma$ (volume fraction~$\phi\approx0.38$)
and $l_{\mathrm{B}} = 3\sigma$ (volume fraction~$\phi\approx0.42$), where $q$ is selected as the wavenumber corresponding to the first peak of the structure factor $S(q)$. See the Methods section for the definitions of $S(q)$ and $F_{s}({q}, t)$.The structural relaxation time~$\tau_\alpha$ is defined as the time when $F_{s}({q}, t)$ decays to $1/e$. We find $\tau_\alpha\approx70\sim100\tau_\text{BD}$. {\bf h, i,} Temporal change in the characteristic timescales of bulk and shear deformation (inverse of strain rate, $\Delta t/|\varepsilon_{\rm bulk}|$ and $\Delta t/|\varepsilon_{\rm shear}|$) for the same PE solution at Bjerrum lengths $l_{\mathrm{B}} = 2\sigma$ ({\bf h}) and  $l_{\mathrm{B}} = 3\sigma$ ({\bf i}). The estimated timescale of the domain deformation $\tau_{\rm def}$ is $5\sim10\tau_{\rm BD}$. These results demonstrate that particle rearrangement characterised by $\tau_\alpha$ is slower than the domain deformation, indicating that network coarsening is controlled by mechanical relaxation. } 
  \label{fig:Fig2}
\end{figure}

In nature, the charges of PEs are generally not perfectly balanced locally during phase separation. 
Moving on to the case of asymmetric charge, we investigate the influence of charge asymmetry between polycations and polyanions by altering the ratio between $N_\mathrm{c}$ and $N_\mathrm{a}$. Figure~\ref{fig:Fig1}b displays the evolution of network-forming domains under $N_\mathrm{c}=50$ and $N_\mathrm{a}=30$. As depicted in Fig.~\ref{fig:Fig2}, the coarsening dynamics of domains slow down in the presence of charge asymmetry, leading to a tendency of dynamic slowing down in the later stages. This deceleration in coarsening holds even when there is only a slight asymmetry in PE charges (e.g., $N_\mathrm{c}=45$, $N_\mathrm{a}=35$). These observations are consistent with prior research~\cite{chen2023charge}, which demonstrates that charge asymmetry significantly impedes the coarsening of individual droplets carrying net charges.

\subsection*{Mechanisms of power-law coarsening and dynamic slowing down}

\begin{figure}[t!]
  \centering
  \includegraphics[width=8.5cm]{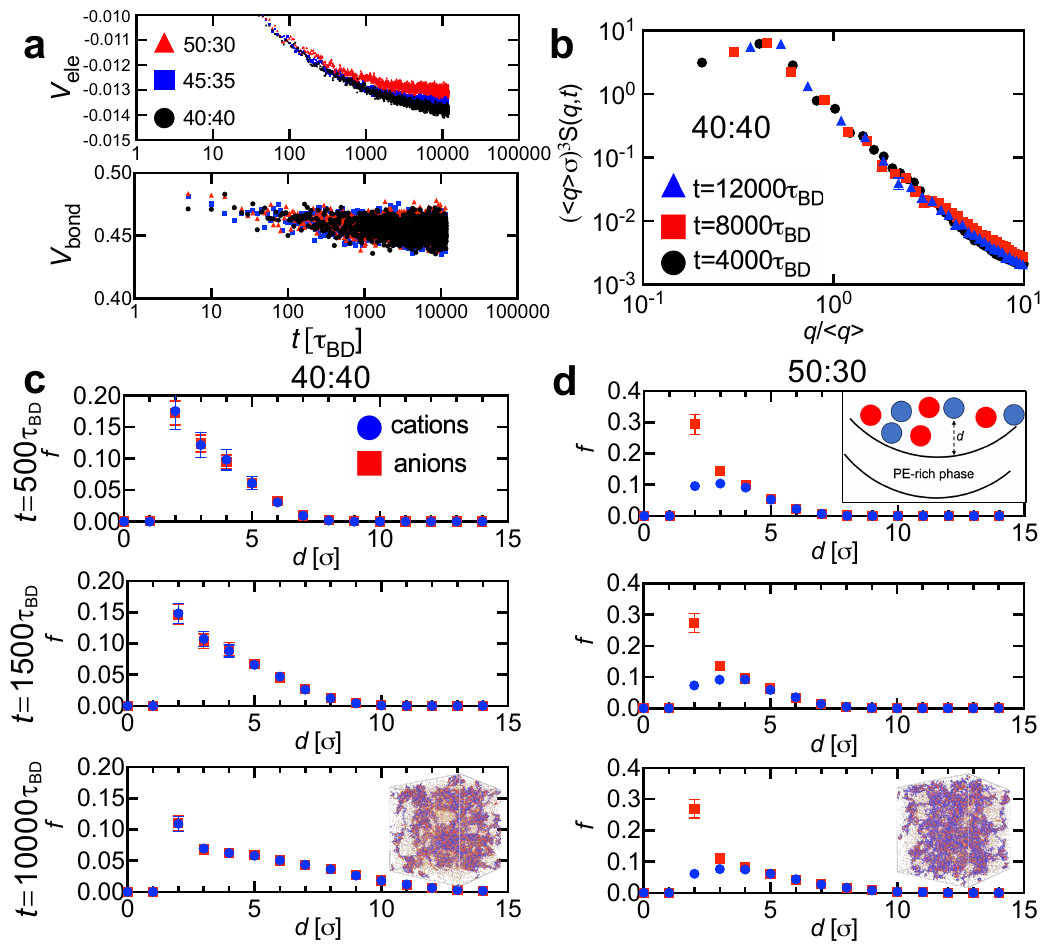}
  \caption{\footnotesize {Characterisation of network-forming phase separation.} {\bf a,} Temporal evolution of  electrostatic energy $V_\mathrm{ele}$, averaged over all particles, for $(N_\mathrm{c}, N_\mathrm{a})=(50,30)$, $(N_\mathrm{c}, N_\mathrm{a})=(45,35)$, and $(N_\mathrm{c}, N_\mathrm{a})=(40,40)$ under $l_{\mathrm{B}} = 2\sigma$ and temporal evolution of bond elastic energy $V_\mathrm{bond}$, averaged over all bonds at $l_{\mathrm{B}} = 2\sigma$. {\bf b,} The scaled structure factors $(\langle q \rangle \sigma)^3S(q,t)$ at different times $t$, which collapse onto a single master curve, consistent with the power-law domain coarsening observed for $(N_\mathrm{c}, N_\mathrm{a})=(40,40)$. {\bf c,} Temporal evolution of counterion distributions (cations and anions) as a function of distance $d$ from the network surface for $(N_\mathrm{c}, N_\mathrm{a})=(40,40)$. {\bf d,} A similar analysis for the charge asymmetry condition $(N_\mathrm{c}, N_\mathrm{a})=(50,30)$. The upper panel of {\bf d} provides a schematic illustrating the distance $d$ of counterions to the network surface (see Methods section for details).} 
  \label{fig:Fig3}
\end{figure}

To provide microscopic insights into the network-forming phase separation, we track the temporal variation of the electrostatic energy averaged over all particles, denoted as $V_\mathrm{ele}$ (Fig.~\ref{fig:Fig3}a). Remarkably, we observe a monotonic decrease in $V_\mathrm{ele}$ throughout the phase separation process, indicating the pivotal role of electrostatic interactions in driving the phase separation. However, upon comparing results between cases of symmetry and asymmetry in PE charge, we notice that $V_\mathrm{ele}$ consistently tends to be larger in the latter case. This discrepancy reflects how the heightened electrostatic repulsion more significantly impedes domain coarsening in the presence of charge asymmetry.

Additionally, the time evolution of the average bond energy, $V_\mathrm{bond}$, remains nearly the same across all three cases (Fig.~\ref{fig:Fig3}a). In our previous study on phase separation dynamics of neutral polymers~\cite{yuan2023mechanical}, we identified the mechanism behind its dynamic slowing down. This mechanism arises from the delay in the relaxation of polymer bonds relative to the rapid deformation of the domains, leading to segment stretching and mechanically hindering domain coarsening. However, our analysis (Fig.~\ref{fig:Fig3}a) rules out this possibility.
This further strengthens the argument that the deceleration in the coarsening process for the case of asymmetric charge is due to electrostatic repulsions, rather than polymer bond-specific viscoelastic effects due to the local chain stretching~\cite{yuan2023mechanical}.
We confirm that the results under $l_{\mathrm{B}} = 3\sigma$ are qualitatively the same (see Supplementary Fig.~S1). 

In the case of charge symmetry, we confirm that the scaled structure factor $(\langle q \rangle \sigma)^3S(q,t)=g(q/\langle q \rangle)$ at different times $t$ ($g(\cdot)$: some function) collapses onto a single master curve (Fig.~\ref{fig:Fig3}b) for $N_\mathrm{c}=N_\mathrm{a}=40$. This supports the self-similar nature of domain shape evolution and the power-law behaviour of domain coarsening. 

Crucially, we observe that under $l_{\mathrm{B}} = 2\sigma\sim3\sigma$, the volume fraction $\phi_\text{d}$ of the dense phase, determined by averaging around the peak in the distribution of the monomers' local volume fraction $\phi=\pi\sigma^3/6V_{\rm vor}$ (where $V_{\rm vor}$ is the Voronoi volume), is approximately $\phi_\text{d} \approx 40$\%. This is notably lower than the $\phi_\text{d} \approx 50$\% found in the LJ attractive neutral polymer systems we previously studied~\cite{yuan2023mechanical}. The looser packing of PE monomers facilitates local bond relaxation without accumulating stretched chains~\cite{yuan2023mechanical}. This is consistent with the difference in shrinking tendency between the collapse of a polymer in a poor solvent~\cite{Kamata2009} and the complexation of a pair composed of a polycation and a polyanion in a good solvent~\cite{chen2022driving,chen2022complexation}.

In the former (VPS) case, strong inter-polymer attractions support stretched polymers, preventing the polymer-rich phase from reaching equilibrium composition (or, volume shrinking), leading to a breakdown of self-similarity and deviation from the power law~\cite{yuan2023mechanical}. In contrast, weak effective inter-polymer attractions due to charge cancellation are insufficient to support stretched polymers in the latter case. Without these stretched chains to support mechanical stress, the polymer-rich phase reaches equilibrium composition more quickly and maintains self-similar, power-law growth.

Nevertheless, the condition $\tau_\alpha \gg \tau_\mathrm{d}$ still holds, unequivocally indicating that the network domain's relaxation is primarily mechanical. Consequently, PE solutions with symmetric charges exhibit the same coarsening power-law as observed in the VPS of colloidal suspentions and solutions of short neutral polymers, characterised by a universal $\langle q \rangle \sim t^{-1/2}$~\cite{tateno2021,yuan2023mechanical}. We speculate that this power-law coarsening may also apply to networks composed of sufficiently long PE chains and hope this will encourage experimentalists to investigate this power-law behaviour.

\subsection*{Scaling derivation for power-law coarsening}

Here, we present a scaling derivation for the power-law coarsening, similar to our previous work~\cite{tateno2021}, to clarify the origin of the growth exponent $\nu = 1/2$. The elastic deformation of the network structure must be accompanied by local changes in composition, $\delta \phi(\mathbf{r})$, relative to the average value $\phi_\text{d}$, such that the local volume deformation is given by $\epsilon(\mathbf{r}) = -\delta \phi(\mathbf{r}) / \phi_\text{d}$. Due to mass conservation, the volume deformation in the dense phase is coupled with solvent transport, resulting in solvent permeation through small gaps between densely packed particles. This scenario aligns with poroelastic theory~\cite{biot1941general}, where the local volume deformation $\epsilon(\mathbf{r}, t)$ satisfies a diffusion-like equation: 
\begin{equation}
\frac{\partial}{\partial t} \epsilon = D_{\mathrm{P}} \nabla^2 \epsilon, \label{eq:poro}
\end{equation} 
where $D_{\mathrm{P}}$ is the poroelastic diffusivity~\cite{biot1941general}. Since the average composition, $\phi_\text{d}$, within the network phase remains nearly constant during coarsening, the elasticity of the dense network phase stays roughly constant over time, allowing us to treat $D_{\mathrm{P}}$ as time-independent. This implies that the mechanical relaxation of the domain network is constrained by the slow solvent transport through the dense phase. Consequently, the characteristic time for domain deformation corresponds to the time required for solvent transport over a characteristic length $\ell$. The self-similar nature of the process suggests the existence of a single characteristic length, $\ell$, justifying its use as the spatial unit in the Laplacian term of Eq.~(\ref{eq:poro}). Thus, the scaling analysis leads to the domain coarsening law $\ell \sim t^{1/2}$.

\subsection*{Temporal evolution of monomer and counterion distributions}

\begin{figure}[t!]
  \centering
  \includegraphics[width=8.5cm]{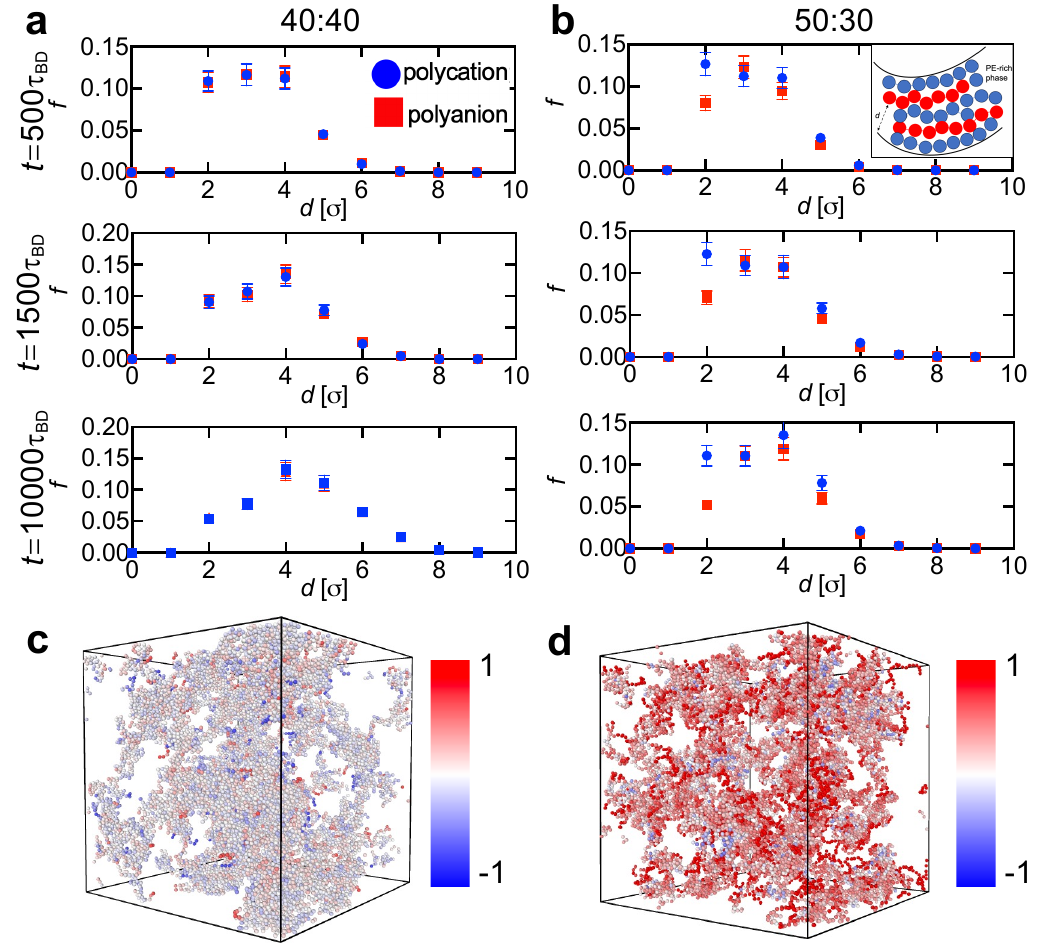}
  \caption{{Distribution of polycations and polyanions during network-forming phase separation.} {\bf a,} Temporal evolution of monomer distributions for polycations and polyanions as a function of the distance $d$ from the network surface for $(N_\mathrm{c}, N_\mathrm{a})=(40,40)$ with a Bjerrum length of $l_{\mathrm{B}} = 2\sigma$. {\bf b,} Similar analysis for the charge asymmetry condition $(N_\mathrm{c}, N_\mathrm{a})=(50,30)$. The upper panel of {\bf b} shows a schematic of  the distance $d$ of monomers to the network surface (see Methods section for details). {\bf c, d,} Configurations at $t=10^4\tau_\mathrm{BD}$ for $(N_\mathrm{c}, N_\mathrm{a})=(40,40)$ and $(N_\mathrm{c}, N_\mathrm{a})=(50,30)$. Particle colour represents the averaged charge, obtained by smearing point charges into a continuous charge density using Gaussian filter $G({\bm r}) = \exp \left(-{{\bm r}^{2}}/{2\sigma^{2}}\right)$ and averaging over grid points within a sphere of radius $3\sigma$.} 
  \label{fig:Fig4}
\end{figure}

The coarsening process induces alterations in the distribution of counterions and monomers around the network, which is typically challenging to measure directly through experiments. We examine the temporal evolution of the counterion (Fig.~\ref{fig:Fig3}c--\ref{fig:Fig4}d) and monomer distribution functions (Fig.~\ref{fig:Fig4}) as a function of their distance $d$ to the network's surface (see the insets of Fig.~\ref{fig:Fig3}d and Fig.~\ref{fig:Fig4}b for schematic representation and Methods section for the definition of $d$). In the case of charge symmetry, we observe identical distributions for both cations and anions (Fig.~\ref{fig:Fig3}c), as well as for the monomers of polycations and polyanions (Fig.~\ref{fig:Fig4}a), consistent with an overall charge-neutralised network (see Fig.~\ref{fig:Fig4}c). Upon phase separation, both cations and anions gradually move away from the network due to the expanding size of void regions where most mobile ions reside, evidenced by the development of tails in the distribution functions (Fig.~\ref{fig:Fig3}c). The observation that ions gradually migrate away from the PE network into the voids reflects the counterion-release entropy in driving phase separation, in addition to the electrostatic attraction. A similar trend is observed in the evolution of monomer distributions (Fig.~\ref{fig:Fig4}a), reflecting the growing domain thickness.

In contrast, when charge asymmetry is present, the anion distributions exhibit a significantly larger peak compared to the cations (Fig.~\ref{fig:Fig3}d). This observation aligns with the fact that the network carries a positive net charge in this scenario. Remarkably, examining the monomer distribution under these conditions reveals that the predominant net charge appears on the surface of the network (Fig.~\ref{fig:Fig4}b; also see Fig.~\ref{fig:Fig4}d). Inside the dense network, polycations and polyanions are distributed nearly identically, as depicted in Fig.~\ref{fig:Fig4}b, aiding in preserving local charge neutrality and reducing electrostatic repulsion.

\subsection*{Impact of volume fraction on coacervate morphology}

\begin{figure}[t!]
  \centering8.5cm
  \includegraphics[width=8.5cm]{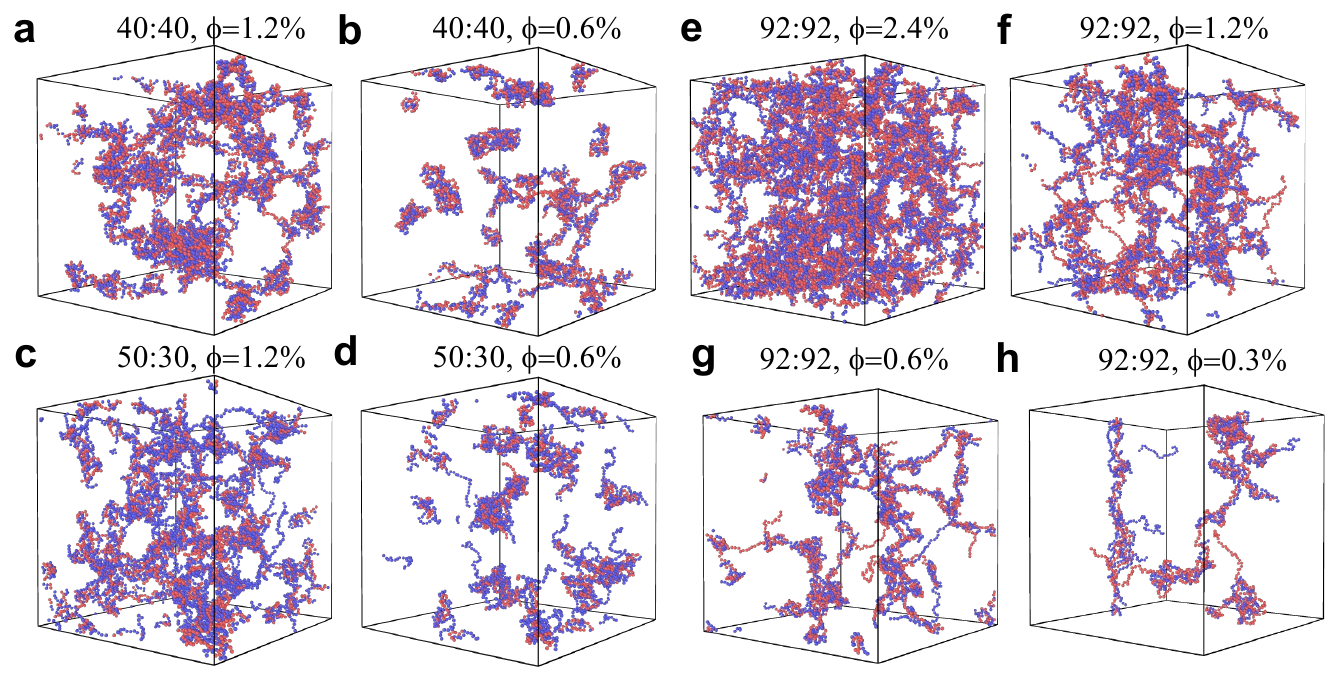}
  \caption{{The phase separation morphology of polycations and polyanions during demixing from an initial homogeneous state under various chain lengths ($N_\mathrm{c}:N_\mathrm{a}$) and volume fraction ($\phi$).} Observation time is $t\approx10^3\tau_\mathrm{BD}$. Specific parameters are indicated at the top of each panel. Bjerrum length is set to $l_{\mathrm{B}} = 2\sigma$.} 
  \label{fig:Fig5}
\end{figure}

We have so far focused on a PE solution with a volume fraction of $\phi = 2.3$\% (similar to the choice in Ref.~\cite{chen2023charge}) and identified a network-forming phase separation morphology starting from a fully mixed state. We anticipate that volume fraction could influence the selection of phase-separation morphology. To investigate this, we maintain the chain lengths $(N_\mathrm{c}, N_\mathrm{a}) = (40,40)$ and $(N_\mathrm{c}, N_\mathrm{a}) = (50,30)$ while reducing the volume fraction $\phi$. Our findings indicated a morphology transition from interconnected networks at $\phi=1.2\%$ (Fig.~\ref{fig:Fig5}a, Fig.~\ref{fig:Fig5}c) to droplets with irregular shape at $\phi = 0.6\%$ (Fig.~\ref{fig:Fig5}b, Fig.~\ref{fig:Fig5}d), suggesting that the crossover volume fraction $\phi^*$ lies between $\phi = 0.6\%$ and $\phi = 1.2\%$.

Notably, we found that the effective volume (calculated as $4/3\pi R_\mathrm{g}^3$, where $R_\mathrm{g}$ is the radius of gyration) occupied by the positive and negative PEs, divided by the box volume, is 1 for $\phi \approx 1\%$. This implies that the effective overlapping of PEs in the initial configuration may be a cause of  network formation. Additionally, longer PEs would exhibit slower conformational relaxation dynamics, which may also favour the selection of network-forming phase separation. Indeed, when the chain lengths were increased to $N_\mathrm{c} = N_\mathrm{a} = 92$, we observed that at the same $\phi = 0.6$\%, where droplets formed for $N_\mathrm{c} = N_\mathrm{a} = 40$ (Fig.~\ref{fig:Fig5}b), a network formed for the longer PE chains (Fig.~\ref{fig:Fig5}g). Consequently, the crossover volume fraction $\phi^*$ is shifted to an even smaller value between $\phi = 0.3\%$ and $\phi = 0.6\%$ (Fig.~\ref{fig:Fig5}e--Fig.~\ref{fig:Fig5}h). The formation of chain linkers within the network is observed (Fig.~\ref{fig:Fig5}g). This can be attributed to the expanded configurations of PE chains in a good solvent, which promote the coarsening of connected clusters. Thus, with sufficiently long PEs, as is often the case in practical applications, we anticipate that network formation will be more likely to occur.

\subsection*{Phase separation dynamics of droplet-like coacervates}
\begin{figure}
  \centering
  \includegraphics[width=8.5cm]{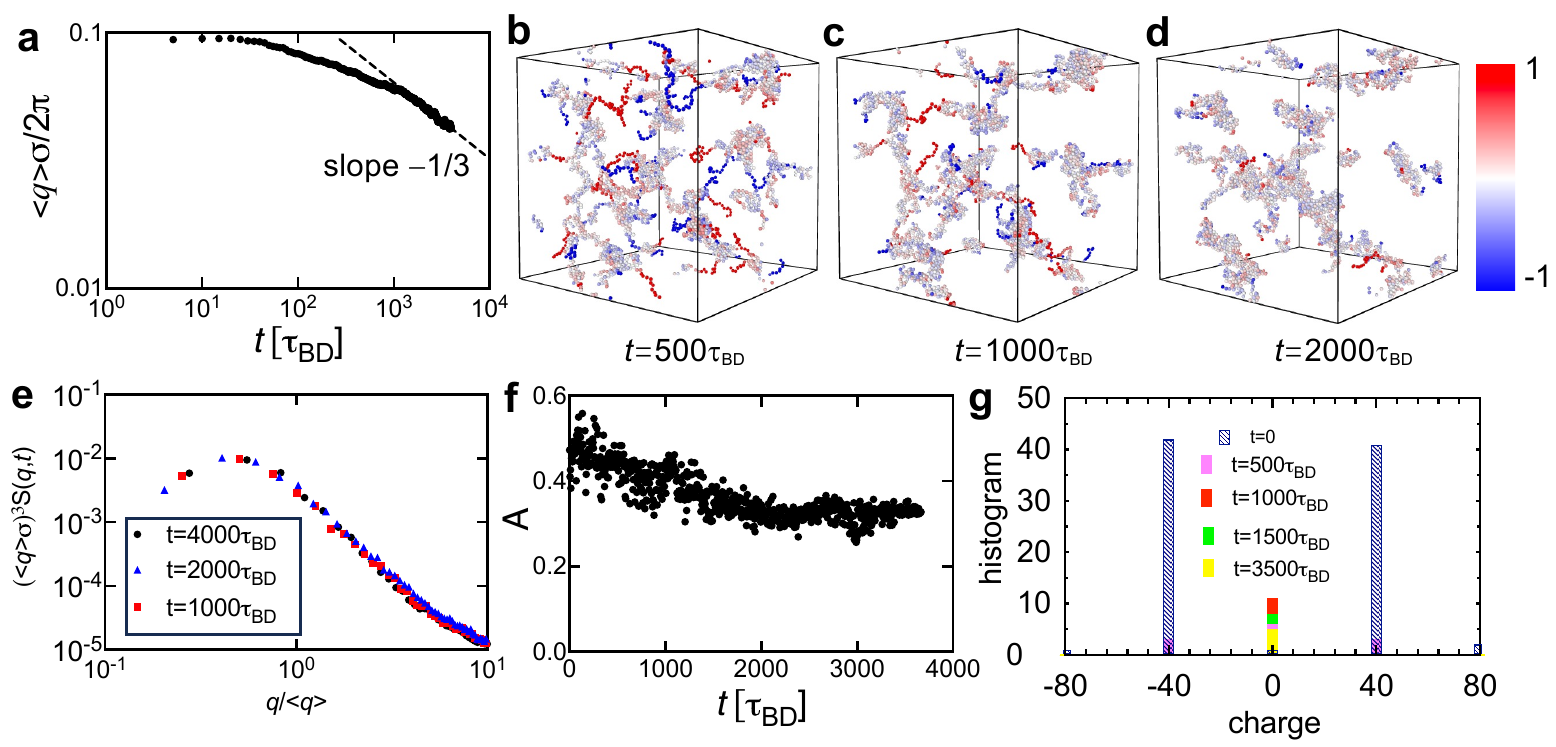}
  \caption{{Droplet-forming phase separation in oppositely charged PEs with chain lengths $(N_\mathrm{c}, N_\mathrm{a}) = (40,40)$ at volume fraction $\phi=0.6\%$ and Bjerrum length $l_{\mathrm{B}} = 2\sigma$.} {\bf a,} Temporal evolution of the characteristic wavenumber $\langle q \rangle$, following power law $\langle q \rangle \sim t^{-1/3}$. {\bf b, c, d,} Simulation snapshots at various times~$t$ (counterions are omitted for clarity).  Particle colour indicates averaged charge, calculated by smearing out point charges into a continuous charge density using Gaussian filter $G({\bm r}) = \exp \left(-{{\bm r}^{2}}/{2\sigma^{2}}\right)$ and averaging over grid points within a sphere of radius $3\sigma$. See Supplemental Movie S3 for visualisations. {\bf e,} Scaled structure factors $(\langle q \rangle \sigma)^3S(q,t)$ at different times $t$ collapsing onto a single master curve, which confirms self-similarity. {\bf f,} Temporal evolution of the degree of asphericity $A$ of individual droplets. {\bf g,} Distribution of net charge carried by individual clusters at various times~$t$, illustrating the heterogeneous charge distribution ({\bf b--d}) in individual clusters despite overall charge neutrality.}
  \label{fig:Fig6}
\end{figure}

The previous valuable study on phase separation dynamics of droplet coacervates assumed a dispersion of microdroplets comprised of a perfectly matched polycation-polyanion pair as the initial state~\cite{chen2023charge}. In our study, we have found that the initiation of phase separation from a fully mixed homogeneous state also results in the formation of isolated, droplet coacervates at sufficiently low volume fractions. For symmetric charges (($N_\mathrm{c}, N_\mathrm{a}) = (40,40)$; see Fig.~\ref{fig:Fig5}b), we observe the temporal evolution of the characteristic wavenumber $\langle q \rangle$ adheres to power law $\langle q \rangle \sim t^{-1/3}$ (Fig.~\ref{fig:Fig6}a), aligning with the findings in Ref.~\cite{chen2023charge}. Additionally, we confirm the maintenance of self-similarity, as the scaled structure factors $(\langle q \rangle \sigma)^3S(q,t)$ at different times $t$ can be collapsed onto a single master curve (Fig.~\ref{fig:Fig6}e). This supports the basis for the power-law growth~\cite{onuki2002phase}.

Here, we note that while droplet-forming phase separation is often mistakenly interpreted as nucleation-growth-type, the phase separation observed in our study clearly follows spinodal decomposition in an unstable state~\cite{Tanaka2000Viscoelastic,tanaka2017phase,shimizu2015novel}. This is evidenced by the rapid exponential growth of the intensity $I(t)=\int d q S(q, t)$ in the early stages. Consequently, the irregular droplets (Fig.~\ref{fig:Fig6}b--\ref{fig:Fig6}d) should be interpreted as true phase-separated structures rather than transient chain aggregates.

We also point out a striking difference in the droplet morphology between phase separation initiated from a homogeneous state and pre-existing neutral nanodroplets~\cite{chen2023charge}. It is evident that the shape of individual droplets in our simulations largely deviates from sphericity (Fig.~\ref{fig:Fig6}b--d), with a degree of asphericity $A\approx0.3$ ($A=0$ for a sphere and $A=1$ for a rod; see Fig.~\ref{fig:Fig6}f where droplet clusters are defined based on a geometric criterion: two monomers are considered to belong to the same droplet if the distance between them is less than $1.5\sigma$). This suggests that PE coacervates exhibit very low interfacial tension due to being in good solvents, especially when compared to neutral polymers in poor solvents. Indeed, previous theoretical and experimental studies have consistently demonstrated that the PE coacervate phase exhibits extremely low interfacial tension~\cite{qin2014interfacial,ali2019characterization,zhang2021interfacial}.
Once small aggregates form, deviations from global charge neutrality lead to inhomogeneous charge distribution, resulting in weak electrostatic interactions between aggregates. This weak effective inter-polymer interaction also contributes to low interfacial tension. Additionally, a single strand extending from a droplet tends to be straight due to intrachain electrostatic repulsions, which are not significantly influenced by interfacial tension.

Indeed, after spatially coarse-graining the point charges, we can clearly see dangling chain threads with a net charge that are exposed outside the droplet (Fig.~\ref{fig:Fig6}b--d), especially in the early stage. This indicates that imperfect matching between polycations and polyanions can occur during the formation of small aggregates, an effect not captured in the previous work that initiates the phase separation from neutral nanodroplets~\cite{chen2023charge}. This also suggests a spatial heterogeneous distribution of charges despite the overall neutrality of these clusters (Fig.~\ref{fig:Fig6}g).

These results therefore reveal several unusual aspects of droplet-like coacervates formed by demixing from a homogeneous initial state: On one hand, they exhibit self-similarity and power-law domain coarsening, akin to classical LLPS~\cite{onuki2002phase} (Fig.~\ref{fig:Fig6}a,~\ref{fig:Fig6}e). On the other hand, starting from a fully mixed state, the network of low-density strands exhibits minimal volume-shrinking tendency, and the transient droplets display nonspherical shapes (Fig.~\ref{fig:Fig6}b--\ref{fig:Fig6}d). This contrasts with the phase separation of neutral polymers in poor solvents: in this case, polymer networks typically exhibit a strong volume-shrinking tendency, leading to a breakdown of self-similarity~\cite{yuan2023mechanical}, while polymer droplets usually show spherical shapes~\cite{Kamata2009,yuan2024hydrodynamic}. Additionally, we observe that in the early stage, the formation of dangling strands with net charges significantly accelerates phase separation. However, as the system evolves, the spatial inhomogeneity of charge distribution within these aggregates decreases, gradually weakening the electrostatic driving forces. This temporal change in effective attractions represents a significant departure from classical LLPS, where the attractions between the components are independent of their spatial distribution and, therefore, remain constant over time.

\subsection*{Impact of solvent quality on phase separation dynamics}

However, in biological systems, short-range non-electrostatic interactions, such as hydrophobic forces or Van der Waals attractions, can occur between monomers. To test the generality of our findings, we introduce short-range attractions between monomers, modelled using Lennard-Jones potentials with typical energy coupling 
$\epsilon=1\sim2k_\mathrm{B}T$ (a reasonable choice for the strength of short-range attractions). We then examine the phase separation dynamics starting from a fully mixed homogeneous state.

\begin{figure}
  \centering
  \includegraphics[width=8.5cm]{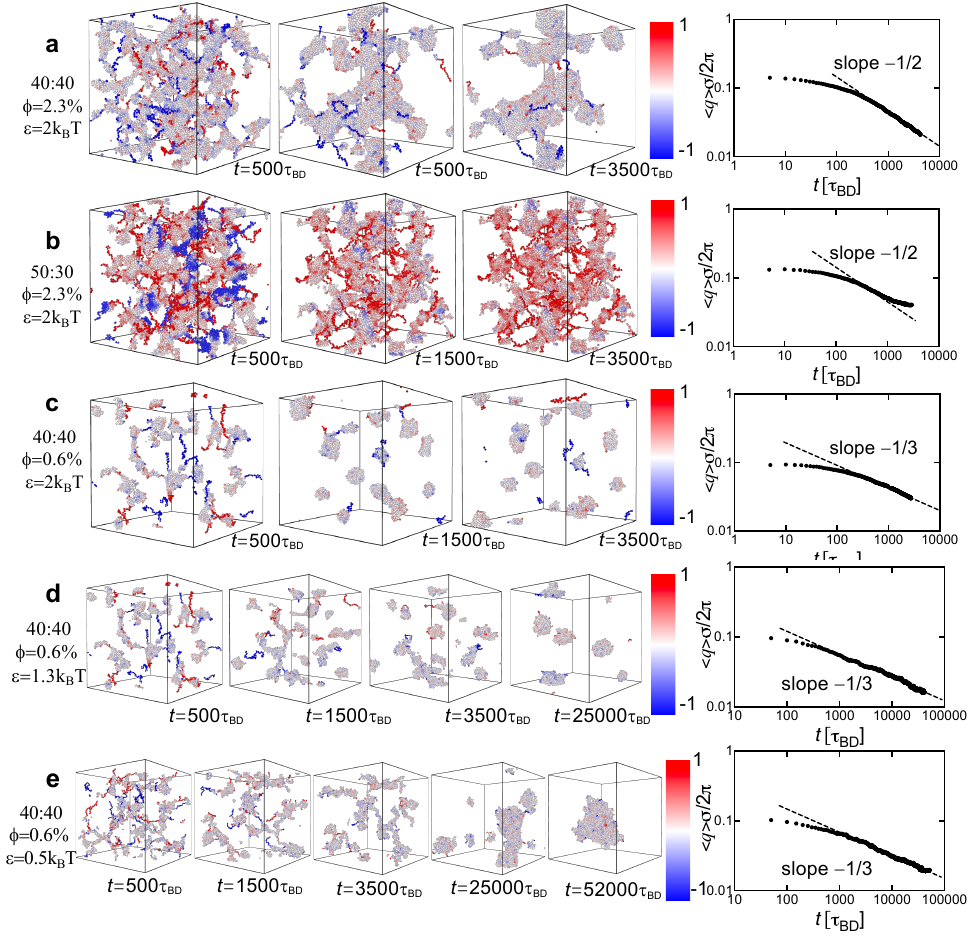}
  \caption{{Network-forming and droplet-forming phase separation dynamics of oppositely charged PEs in poor solvents with Bjerrum length $l_{\mathrm{B}} = 2\sigma$.}  {\bf a,} Parameters: chain lengths $(N_\mathrm{c}, N_\mathrm{a}) = (40,40)$ and volume fraction $\phi=2.3\%$. Snapshots at various times~$t$ (counterions omitted for clarity) and the temporal evolution of the characteristic wavenumber $\langle q \rangle$ (dashed line with slope $-1/2$). Particle colour in snapshots indicates averaged charge (see the caption of Fig.~\ref{fig:Fig5} for details). {\bf b,} Same analysis as in panel {\bf a} with parameters $(N_\mathrm{c}, N_\mathrm{a}) = (50,30)$ and $\phi=2.3\%$. Dashed line has a slope of $-1/2$. {\bf c-- e,} Same analysis as in panel {\bf a} with parameters $(N_\mathrm{c}, N_\mathrm{a}) = (40,40)$ and $\phi=0.6\%$. Dashed line has a slope of $-1/3$. Short-range attractions between monomers are modelled by Lennard-Jones potentials with coupling $\epsilon=2k_\mathrm{B}T$ in {\bf a}--{\bf c}, $\epsilon=1.3k_\mathrm{B}T$ in {\bf d}, and $\epsilon=0.5k_\mathrm{B}T$ in {\bf e}. See Supplemental Movie S4 and S5 for visualisations of {\bf d} and {\bf e}.
  }
  \label{fig:Fig7}
\end{figure}

We observe that network-forming phase separation of oppositely charged PEs follows a $1/2$ power-law in the case of symmetric charge $(N_\mathrm{c}, N_\mathrm{a}) = (40,40)$ (Fig.~\ref{fig:Fig7}a) and exhibits dynamical slowing down in the case of asymmetric charge $(N_\mathrm{c}, N_\mathrm{a}) = (50,30)$ (Fig.~\ref{fig:Fig7}b), confirming the robustness of our findings under poor solvent conditions. Previously, we found that neutral polymers of similar chain lengths also exhibit a $1/2$ power-law under the same condition of $\epsilon=2k_\mathrm{B}T$~\cite{yuan2023mechanical}. This suggests that the inclusion of short-range attractions causes the behaviour of PE networks to resemble that of neutral polymer networks more closely, further supporting the idea that the slowing down in domain coarsening is purely due to electrostatic effects in the case of asymmetric charge $(N_\mathrm{c}, N_\mathrm{a}) = (50,30)$. Overall, these results confirm the generality of our findings on the phase separation dynamics of network-like coacervates in both good and poor solvents.

For droplet-like coacervates in poor and slightly poor solvents (Fig.~\ref{fig:Fig7}c--\ref{fig:Fig7}e), we again observe a $1/3$ power-law in the case of symmetric charge $(N_\mathrm{c}, N_\mathrm{a}) = (40,40)$, consistent with the domain coarsening law under good solvent conditions (Fig.~\ref{fig:Fig5}a) and previous simulations~\cite{chen2023charge}. In all the cases, regardless of the strength of hydrophobic attraction ($\epsilon=2k_\mathrm{B}T$ in Fig.~\ref{fig:Fig7}c, $\epsilon=1.3k_\mathrm{B}T$ in Fig.~\ref{fig:Fig7}d, and $\epsilon=0.5k_\mathrm{B}T$ in Fig.~\ref{fig:Fig7}e), the early stage features imperfect matching between polycations and polyanions, along with the formation of dangling chain threads with net charges, similar to what is observed in good solvent conditions (Fig.~\ref{fig:Fig5}). The overall-neutral coacervate droplets display irregular shapes for an extended period in slightly poor solvents characterised by $\epsilon=1.3k_\mathrm{B}T$ (Fig.~\ref{fig:Fig7}d) and eventually evolve into spherical shapes at a later stage of $t\approx2.5\times10^4\tau_\mathrm{BD}$ (Fig.~\ref{fig:Fig7}d). Droplets become spherical much more quickly under stronger short-range attraction of $\epsilon=2k_\mathrm{B}T$ (Fig.~\ref{fig:Fig7}c), but remain irregularly shaped within our simulation timescale $t\approx5.2\times10^4\tau_\mathrm{BD}$ under $\epsilon=0.5k_\mathrm{B}T$ (Fig.~\ref{fig:Fig7}e).  These findings indicate that coacervates are inherently irregularly shaped during phase separation, with the transformation into spherical morphologies proceeding slowly in good and slightly poor solvents due to the weak interfacial tension~\cite{qin2014interfacial,ali2019characterization,zhang2021interfacial}.

\subsection*{Strengths and limitations of our coarse-grained model}

Finally, we discuss the validity and limitations of our model in relation to our findings. Our coarse-grained PE model lacks a detailed molecular description of water, thereby neglecting crucial factors such as the structural arrangement of solvent molecules near ions~\cite{van2016water} and the emergence of polarisation effects. These phenomena are complex and significant in aqueous electrolyte solutions. We postulate that the explicit inclusion of water could influence the effective electrostatic attraction between oppositely charged monomers of polycations and polyanions, akin to modulating ion-pair binding strength~\cite{van2016water}. For example, the permittivity of an electrolyte solution can be reduced by the orientational ordering of water~\cite{van2016water}, which may enhance electrostatic coupling at short ion-ion distances, thereby facilitating the occurrence of network-forming VPS~\cite{tanaka1993,Koyama2007}. Importantly, our research demonstrates the emergence of a growth exponent of $\nu=1/2$ under charge symmetry conditions and a dynamic slowing down mechanism of electrostatic origin for charge asymmetry across a wide range of Bjerrum lengths $l_\text{B}$ ranging from $1.1\sigma$ to $3\sigma$, which quantifies the electrostatic strength. Notably, the mechanisms driving our findings are inherently physical and valid on a significantly larger length scale than the typical size of solvent molecules. Therefore, we anticipate that explicitly incorporating water will not markedly alter our conclusions. 

As in our earlier studies on coarse-grained polymer models~\cite{Kamata2009,yuan2023mechanical,yuan2024impactprotein,yuan2024hydrodynamic}, we treat the solvent as a continuous medium. This approach allows for the efficient incorporation of hydrodynamic degrees of freedom using the FPD method~\cite{tanaka2000}, ensuring momentum conservation and solvent incompressibility, which are crucial for modelling particle dynamics in a solvent. Although using explicit water can provide a more accurate description of hydrodynamic couplings between monomers, conducting simulations that incorporate explicit water is technically impractical for analysing phase separation dynamics. Indeed, coarse-grained models are essential and widely utilised for investigating phase separation in PE systems~\cite{shakya2020role,chen2023charge,chen2022driving,chen2022complexation}. 

In this work, we adopt a typical size of hydrated ions,  $\sigma=0.72$~nm. Thus, the length scale of the domain network is around 10~nm, with a simulation box size $L\approx50$~nm. Time is scaled using Brownian time~$\tau_\mathrm{BD}$, which can be derived from the Stokes-Einstein relation, $\tau_\mathrm{BD} = 3 \pi \sigma^3 \eta / k_{\mathrm B}T$, where $\eta$ represents the viscosity of the solvent. For water at room temperature ($\eta \approx10^{-3}$~Pas), $\tau_\mathrm{BD} \approx 0.85$~ns. Consequently, our simulation timescale ranges from 10 to 100 microseconds. 
Achieving such length and time scales in atomistic simulations, necessary for studying phase separation, is highly challenging, even with fixed-charge water models. 

For example, a recent atomistic simulation study utilising atomistic TIP3P water similarly reported a porous network structure formed by oppositely charged PEs~\cite{eneh2024solid}, supporting that the network structures observed in our work are not artefacts resulting from the use of FPD~\cite{tanaka2000}. However, these atomistic simulations~\cite{eneh2024solid} were constrained by using $3\sim4$ times fewer chains and much shorter timescale ($\sim200$~ns), thus unable to provide insights into long-time phase separation dynamics, such as power-law coarsening behaviour. 

Furthermore, in this work, we also observe $\nu = 1/3$ for droplet-forming phase separation under a charge-matched condition, consistent with the classical growth exponent in liquid-liquid phase separation. This implies that details at length scales smaller than the monomer size may not significantly affect the power-law scaling. This situation parallels the coarsening dynamics of ordinary phase separation, which exhibits universality irrespective of the microscopic details~\cite{onuki2002phase}. While the prefactor may be influenced by these microscopic details, the growth exponents, or scaling relations, are expected to remain independent of such details. Exploring the intricate impacts of water on the microscopic features of coacervates, such as hydration effects and local water ordering, is a crucial avenue for future research~\cite{sing2020recent}.

\section*{Discussion}

The primary novelty of this work lies in expanding the traditional view of coacervates as spherical droplets, with droplet growth governed by ordinary liquid-liquid phase separation mechanisms. First, we demonstrate that starting from a homogeneously mixed state, coacervates form space-spanning networks instead of droplets, even at relatively low PE concentrations. This network formation arises from the slow relaxation dynamics of the dense phase compared to the fast domain deformation speed during phase separation. This dynamical asymmetry triggers VPS~\cite{Tanaka2000Viscoelastic}, a phenomenon commonly observed in polymer solutions~\cite{Tanaka1996}, colloidal suspensions~\cite{tateno2021}, and globular proteins~\cite{Tanaka2005Viscoelastic} undergoing demixing. In particular, the power-law coarsening characterised by the growth exponent $\nu=1/2$ for charge-matched systems aligns with VPS behaviour in neutral low-molecular-weight polymer solutions~\cite{yuan2023mechanical} and colloidal suspensions~\cite{tateno2021}. 
Unlike neutral long polymers in poor solvents, self-similarity is consistently maintained due to the weaker interfacial tension and the reduced collapsing tendencies of PE pairs in good solvents. These findings collectively highlight the generality of VPS and suggest its relevance to coacervate phase separation.

Here, we note that this growth exponent $\nu=1/2$ observed in our study contrasts with $\nu=1/3$ seen in droplet-forming phase separation of oppositely charged PE mixtures initiated from nanodroplets~\cite{chen2023charge}. Beyond the difference in growth exponents, we emphasise the fundamental distinction in the coacervate morphologies: space-spanning networks versus droplets. These findings suggest that the commonly observed droplet-shaped coacervates may, in part, arise from imperfect mixing at the initial stage. In contrast, homogeneous mixing activates the viscoelastic effects of the PE-rich phase during phase separation, leading to network formation, i.e., triggering the VPS.
Together with the results from Chen and Wang~\cite{chen2023charge}, our study underscores the critical role of initial configurations --- specifically the protocol used to initiate phase separation --- in determining domain morphology and coarsening dynamics. This provides an interesting and useful way to activate or deactivate viscoelastic effects by adjusting the initial conditions. Indeed, experiments suggest that networks composed of oppositely charged PEs formed through desalting exhibit greater mechanical performances compared to other protocols such as solvent drying~\cite{murakawa2019polyelectrolyte}. Our simulation setup mimics the former scenario.

Second, in droplet-forming phase separation, we show that coacervate droplets exhibit irregular, long-lasting shapes and transition into spherical forms only slowly due to weak interfacial tension, particularly in good solvents. The persistence of these non-spherical droplets deviates from the conventional expectation of perfectly spherical morphologies, revealing an intriguing and underexplored characteristic of PE coacervates.

Although simulating charged PEs in molecular dynamics has become commonplace, incorporating both long-range electrostatic and HI significantly increases computational expenses. Accordingly, free-draining Brownian dynamics or Langevin dynamics simulations have been widely employed in studying the dynamics of coacervate formation. To underscore the impact of HI, we also investigate network domain coarsening using Brownian dynamics simulations without HI and observe a growth exponent of $\nu=1/3$, highlighting the essential roles of HI in the network coarsening of oppositely charged PEs. This demonstrates that including hydrodynamic degrees of freedom is essential for adequately modelling the non-equilibrium dynamics of PEs undergoing demixing.

While numerical simulations provide critical microscopic insights, it is important to recognise that the accessible time and length scales are significantly smaller than those achievable in experiments. This limitation poses challenges in investigating the macroscopic time and length scales relevant to phase separation. 
We hope that our work inspires future studies to delve deeper into the viscoelastic effects~\cite{Tanaka2000Viscoelastic, spruijt2013linear} on the domain coarsening dynamics of coacervates, extending beyond the traditional frameworks of liquid-liquid phase separation.

\section*{Conclusion and outlook}

In summary, we use fluid particle dynamics simulations~\cite{tanaka2000,yuan2024hydrodynamic} incorporating  electrostatics, hydrodynamic interactions (HI), and explicit ions to systematically investigate domain coarsening of oppositely charged PEs in semi-dilute solutions. We challenge the traditional view that coacervates are predominantly  spherical droplets and that domain growth adheres to the physical mechanisms of classical liquid-liquid phase separation. Starting from a thoroughly mixed state, oppositely charged PEs spontaneously form a percolated network, even in semi-dilute solutions. Notably, we observe an unconventional growth exponent of $\nu=1/2$ under charge symmetry conditions across a broad range of Bjerrum lengths, reminiscent of the network-forming viscoelastic phase separation of neutral low-molecular-weight polymers without dynamic slowing down~\cite{yuan2023mechanical} and colloidal suspensions~\cite{tateno2021}.

Additionally, we observe a morphology transition from an interconnected network structure at a volume fraction of $\phi=1.2$\% to a typical droplet-like morphology  at $\phi=0.6$\% for a degree of polymerisation $N \sim 50$. This indicates that a volume fraction above approximately 1\% is necessary to induce network formation. This crossover volume fraction can be further reduced with longer PEs due to their slower relaxation dynamics. This presents a novel approach to forming network structures at extremely low polymer volume fractions, with promising implications for the development of network and porous materials in practical applications.
We also reveal that the observed droplet-like coacervates in good solvents are irregularly shaped, which differs from the typical spherical droplets seen in the usual LLPS. This also explains why percolation can be easily achieved at low concentrations of PEs.

While focusing on coacervate formation in good solvents, we confirm that key behaviours of network phase separation persist under poor solvent conditions. These include power-law scaling for symmetric charge interactions between polycations and polyanions, dynamical slowing down for asymmetric charge interactions, and spatial charge inhomogeneity in irregularly shaped droplets during the early stages of phase separation. As solvent quality decreases, coacervate droplets more readily adopt spherical shapes, whereas in good and slightly poor solvents, droplets retain irregular morphologies for longer periods. Our findings indicate that oppositely charged polyelectrolyte condensates are initially irregularly shaped, with a gradual transition to spherical forms that occurs more slowly in good solvents. This behaviour may explain experimental observations of nonspherical droplets~\cite{lu2020multiphase}, likely resulting from weak interfacial tension due to the absence of short-range, non-electrostatic attractions in such conditions, which inhibits the rapid formation of spherical shapes.

As the degree of charge asymmetry increases, the coarsening process slows down due to the net charge buildup at the surface of the interconnected network, leading to a tendency toward dynamic slowing down of electrostatic origin. Therefore, the stability of these networks can be regulated by varying the charge asymmetry. This charge-dependent regulation of network stability has significant implications for both biological and industrial applications. For instance, stable, mesh-like biological condensates capable of supporting mechanical forces are essential for mechanics-related processes like cell division. Similarly, such mechanisms can be harnessed for the design of stable porous materials.

While droplet-like biological condensates are frequently observed in biological cells, recent studies have reported network-like morphologies~\cite{Woodruff2015,berry2018physical,Ma2021,tanaka2022Viscoelastic}, such as centrosome assemblies~\cite{Woodruff2015}, protein granules~\cite{An2021}, and TIS granule network~\cite{Ma2021}. However, the mechanisms underlying their formation have remained elusive. Recent research has connected these network-like condensates to viscoelastic phase separation (VPS) observed in polymer solutions~\cite{Ma2021,tanaka2022Viscoelastic}. It has been proposed that the selective incorporation of polymeric components, such as DNA and RNA, leads to dynamic asymmetry, triggering the formation of networks stabilised by inter-chain cross-linking~\cite{tanaka2022Viscoelastic}.

Our findings highlight the critical role of electrostatic interactions in this process. Specifically, charge asymmetry can induce transient networks and dynamic slowing down, providing a physical basis for the formation of stable, network-like condensates. This perspective advances our understanding of the mechanisms underlying neurodegenerative diseases, where electrostatic interactions play a pivotal role. For example, the microtubule-binding protein Tau undergoes liquid-liquid phase separation through electrostatically driven complex coacervation under physiological conditions~\cite{lin2020electrostatically, najafi2021liquid, gracia2022molecular}. 

Beyond biological systems, these insights can guide the design of porous structures for industrial applications. Our simulation methodology provides a versatile tool for investigating electro-hydrodynamic couplings in various solvated soft materials, including hydrogels and membrane filters, where network structures arise from the phase separation of oppositely charged polyelectrolyte (PE) systems~\cite{Wu2010, murakawa2019polyelectrolyte, sadman2019versatile, durmaz2020polyelectrolyte, baig2020sustainable}.
For example, Murakawa et al. reported the formation of stable porous networks during the desalting of a homogeneous mixture of anionic PNaSS and cationic PDADMAC, initially placed in high ionic strength solutions~\cite{murakawa2019polyelectrolyte}. Similarly, Durmaz et al. employed the desalting protocol to initiate the phase separation dynamics of highly charged PEs, anionic PSS and cationic PDADMAC, a model system forming coacervates, and observed the formation of stable porous network materials~\cite{durmaz2020polyelectrolyte}. We note that these polymers are considered `typical' examples of strongly charged, flexible, or semi-flexible PE chains, indicating that the formation of stable networks through the mechanism described here is broadly applicable to coacervates.

Finally, our results align with experimental observations showing a substantial increase in viscosity and shear modulus upon adding oppositely charged PEs to solutions~\cite{spruijt2013linear}. Together, these insights emphasise the versatility of electrostatically driven polymer phase separation in forming stable, functional networks with applications ranging from biological systems to industrial materials.

\vspace{1cm}
\noindent
{\bf Acknowledgements} This work was supported by the Grant-in-Aid for Specially Promoted Research (JSPS KAKENHI Grant No. JP20H05619) from the Japan Society for the Promotion of Science (JSPS).
The authors thank the Supercomputer Center, the Institute for Solid State Physics, the University of Tokyo for resources.  
\vspace{2cm}

\setcounter{figure}{0}
\renewcommand{\thefigure}{S\arabic{figure}}
\renewcommand{\thetable}{S\arabic{table}}

\begin{figure}
  \centering
  \includegraphics[width=8cm]{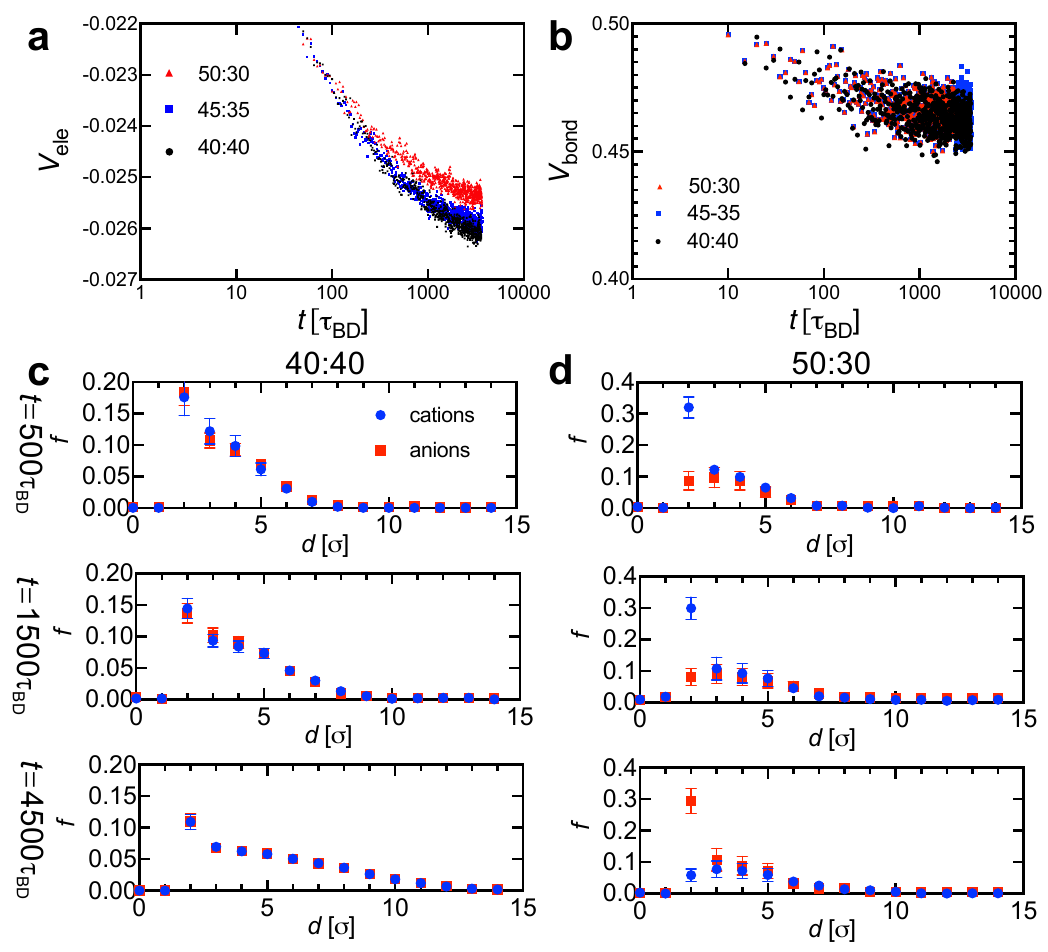}
  \caption{{Characterisation of network-forming phase separation under $l_{\mathrm{B}} = 3\sigma$.} 
{\bf a,} Temporal evolution of electrostatic energy $V_\mathrm{ele}$ averaged over all particles for different chain length settings: $(N_\mathrm{c}, N_\mathrm{a})=(50,30)$, $(N_\mathrm{c}, N_\mathrm{a})=(45,35)$, and $(N_\mathrm{c}, N_\mathrm{a})=(40,40)$, under $l_{\mathrm{B}} = 3\sigma$. {\bf b,} Temporal change of bond elastic energy $V_\mathrm{bond}$ averaged over all bonds under $l_{\mathrm{B}} = 3\sigma$. {\bf c,} Temporal evolution of cations and anions distribution as a function of the distance d to the network surface for the charge-balanced condition $(N_\mathrm{c}, N_\mathrm{a})=(40,40)$. {\bf d,} Similar analysis as in {\bf c} for the charge asymmetry condition $(N_\mathrm{c}, N_\mathrm{a})=(50,30)$.} 
  \label{fig:FigS1}
\end{figure}

\begin{figure}
  \centering
  \includegraphics[width=8cm]{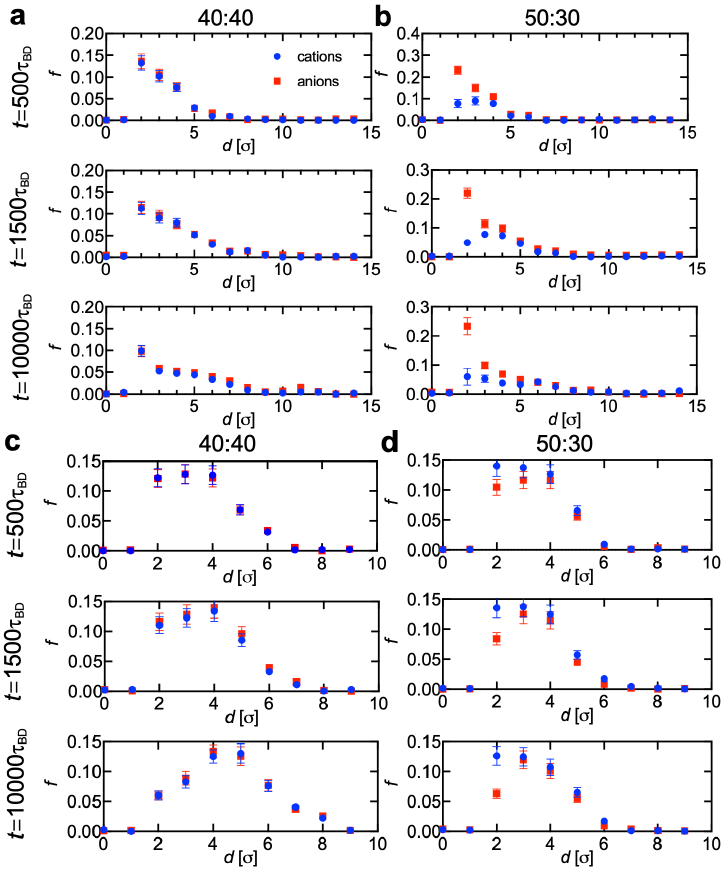}
  \caption{{Characterisation of ion and monomer distribution around the network surface under $l_\text{B}=2\sigma$.}
  Here we employ the same approach as described in the main text, except using a different cutoff $\rho_{\text {th}}=0.4$ for defining the surface. {\bf a,} Temporal evolution of cation and anion distribution as a function of the distance $d$ from the network surface for the charge-balanced condition $(N_\mathrm{c}, N_\mathrm{a})=(40,40)$. {\bf b,} Similar analysis as in {\bf a} for the charge asymmetry condition $(N_\mathrm{c}, N_\mathrm{a})=(50,30)$. {\bf c,} Temporal evolution of polycation and polyanion distribution as a function of the distance $d$ from the network surface for the charge-balanced condition $(N_\mathrm{c}, N_\mathrm{a})=(40,40)$. {\bf d,} Similar analysis as in {\bf c} for the charge asymmetry condition $(N_\mathrm{c}, N_\mathrm{a})=(50,30)$. These results are qualitatively consistent with those in the main text.}
  \label{fig:FigS2}
\end{figure}

\section*{Methods}

\subsection*{Polyelectrolyte modelling}

We employ a coarse-grained model similar to that used in our recent work on the collapse dynamics of a PE chain~\cite{yuan2024hydrodynamic}. The salt-free solution comprises cationic PEs (with monomer charge $q_\mathrm{c}=e$) and anionic PEs (with monomer charge $q_\mathrm{a}=-e$), along with monovalent counterions, all contained in a cubic periodic simulation box of size $L$ (see Fig.~\ref{fig:Fig1}). The cationic and anionic PEs are represented as bead-spring chains~\cite{stevens95}, with each chain consisting of $N_\mathrm{c}$ and $N_\mathrm{a}$ monomers, respectively. The ions and monomers are modelled as spheres with a diameter of $\sigma$, interacting via a purely repulsive shifted-truncated Lennard-Jones (LJ) potential with a coupling constant $\varepsilon_{\mathrm{LJ}}=k_{\mathrm{B}}T$, where $k_{\mathrm{B}}$ is Boltzmann's constant and $T$ is the absolute temperature. The bonds between neighbouring monomers are described by a harmonic potential $V_{\mathrm{bond}}=(K/2)(r-r_0)^2$, with a spring constant $K=400k_{\mathrm{B}}T/\sigma^2$ and a bond length $r_0=2^{1/6}\sigma$. This harmonic potential ensures that the bond lengths remain close to equilibrium values. As customary for coarse-grained PE simulations, we consider the Bjerrum length $l_{\mathrm{B}} = e^{2} / (4\pi\varepsilon_{\mathrm{sol}}k_{\mathrm{B}}T) = 1\sim3\sigma$, where $\varepsilon_{\mathrm{sol}}$ denotes the solvent permittivity.

Each simulation includes 184 cationic and 184 anionic chains within a box of size $L=69.2\sigma$. Consistent with prior work~\cite{chen2023charge}, we vary the ratio between $N_\mathrm{c}$ and $N_\mathrm{a}$ to investigate the effect of charge asymmetry while maintaining $N_\mathrm{c}+N_\mathrm{a}=80$. The volume fraction of monomers is $\phi\approx2.3$\%, corresponding to a semi-dilute solution. 

The system is initially equilibrated using purely repulsive Weeks-Chandler-Andersen (WCA) interactions, forming a thoroughly mixed, homogeneous state. Subsequently, electrostatic interactions are activated to initiate the phase separation. The computation of electrostatic interactions employs the Ewald summation method, ensuring a relative force accuracy of $\epsilon_\mathrm{err}=10^{-3}$.

\subsection*{Fluid particle dynamics}

To account for hydrodynamic interactions (HI), we employ the fluid particle dynamics (FPD) method~\cite{tanaka2000,yuan2022Impact,yuan2024hydrodynamic}, which directly solves the Navier--Stokes (NS) equation.
In the framework of the FPD method, each particle $n$ is represented as a viscous fluid particle, given by the following equation:
\begin{equation}
\psi_{n}({\bm r})=\frac{1}{2}\left\{\tanh
[\left(a-\left|{\bm r}-{\bm R}_{n}\right|\right) /
\xi]+1\right\}\;,
\end{equation}
where a is the particle radius, $\xi$ is
its interface thickness, and ${\bm R}_{n}$ is the position vector of particle $n$, enabling us to obtain a smooth viscosity field $\eta({\bm r})=\eta_{\rm s}+\left(\eta_{\rm p}-\eta_{\rm s}\right)
\sum_{n=1}^{N} \psi_{n}({\bm r})$, where $\eta_{\rm s}$ and
$\eta_{\rm p}$ are the viscosities of fluid and particles, respectively, and
$N$ is the total number of all the particles, including PE monomers and their counterions. 
We calculate the flow field ${\bm v}$ by solving the Navier--Stokes (NS) equation
$\rho\left(\frac{\partial}{\partial t}+\boldsymbol{v} \cdot \nabla\right)
\boldsymbol{v}=\boldsymbol{f}+\nabla
\cdot\left(\boldsymbol{\sigma}+\boldsymbol{\sigma}^{\mathrm{R}}\right)$, where $\rho$ is the constant fluid density, ${\bm f}=\sum_{n}
{{\bm F}_{n} \psi_{n}({\bm r})}/{\int
  \psi_{n}\left({\bm r}^{\prime}\right) d
  {\bm r}^{\prime}}$ is the force field by smearing out the
interaction force ${\bm F}_n$ on the particles. 
$\boldsymbol{\sigma}=\eta(\mathbf{r})$ $\left(\nabla
\mathbf{v}+(\nabla \mathbf{v})^{\mathsf{T}}\right)-p \mathbf{I}$ represents the internal stress term, where $\mathbf{I}$ is the unit tensor and $p$ is the pressure which is determined to
satisfy the incompressible condition $\nabla\cdot{\bm v}=0$. 
The random fluctuating stress field ${\bm \sigma}^{\mathrm{R}}$ are introduced to satisfy the
fluctuation-dissipation relation
$\left\langle{\bm \sigma}^{\mathrm{R}}\right\rangle={\bm 0}$
and $\left\langle\sigma_{i j}^{\mathrm{R}}({\bm r}, t) \sigma_{k
  l}^{\mathrm{R}}\left({\bm r}^{\prime},
t^{\prime}\right)\right\rangle=2 \eta({\bm r}) k_{\rm B}
T\left(\delta_{i k} \delta_{j l}+\delta_{i l} \delta_{j k}\right)
\delta\left(t-t^{\prime}\right)$ where $\delta$ is the Dirac delta function. 
We update the flow field~${\bm v}$ of the NS equation by using the Marker-and-Cell (MAC)
method with a staggered lattice under periodic boundary conditions.

We set the length unit as the lattice size $l_0$, the time unit as $\tau_{0}=\rho
l_{0}^{2} / \eta_{\mathrm{s}}$, $a=3.2l_0$, $\xi=l_0$, particle diameter $\sigma=2a+\xi=7.4l_0$,
$\eta_{\rm p}=2\eta_{\rm s}$, and time step $\Delta t=2.5\times10^{-2}\tau_{0}$.
Following previous work~\cite{tateno2019}, we use
$\beta=1/(k_{\rm B}T)=0.07\rho/(l_0\eta_{\rm s}^2)$
to achieve diffusive dynamics within a reasonable simulation time. 
The simulation time is normalised using the Brownian time of a single particle, defined as $\tau_{\rm BD}=\sigma_{\rm H}^{2} / 24 D$, where $D=k_{\rm B} T /\left(3 \pi \sigma_{\rm H} \eta_{\mathrm{s}}
\lambda\right)$ represents the diffusion constant, $\sigma_{\rm H}=8.14l_0$ is the hydrodynamic diameter~\cite{tateno2019}, and $\lambda=\int
\psi({\bm r}) d {\bm r} / \int \psi({\bm r})^{2} d
{\bm r}$ serves as a correction factor (with $\lambda{\approx}1.69$ for
$\xi/a=1/3.2$). This normalisation results in a ratio of $\tau_{0}/\tau_{\rm BD}{\approx}0.04$.

\subsection*{Brownian dynamics}

To make comparisons, we also conduct free-draining Brownian dynamics (BD) 
simulations using the same PE model. We update
the system configuration by integrating the equation $d
{\bm R}_{n}(t) / d
t={\bm V}_{n}(t)=\zeta^{-1}\left({\bm F}_{n}+{\bm F}_{n}^{R}\right)$, 
where ${\bm R}_{n}(t)$ is the position vector of the particle with the index $n$, ${\bm V}_{n}(t)$ is the particle velocity,
${\bm F}_{n}$ is the interaction force, ${\bm F}_{n}^{R}$ is the random force which satisfies the fluctuation-dissipation relation
$\left\langle{\bm F}_{n}^{\mathrm{R}}\right\rangle={\bm 0}$ and $\left\langle {F}_{n i}^{\mathrm{R}}(t) F_{m
  j}^{\mathrm{R}}\left(t^{\prime}\right)\right\rangle=2 k_{\rm B}T \zeta \delta_{n m} \delta_{i j} \delta\left(t-t^{\prime}\right)$,
where $\zeta$ is the friction constant.
We adopt the monomer diameter $\sigma$ as the length unit, with the time unit defined as $\tau_0=\sigma^{2} / D$, where $D=k_{\rm B} T / \zeta$ represents the diffusion constant of a single particle. The Brownian time in BD simulations is $\tau_{\rm BD}=\sigma^2 /24D$, resulting in $\tau_0=24\tau_{\rm BD}$. To update the system configurations, a time step of $\Delta t=10^{-5}\tau_0$ is utilised.

\subsection*{Structure factor $S(q,t)$ and characteristic wavenumber $\langle q \rangle$} 

To quantify the domain coarsening dynamics, we calculate the structure factor $S(q,t)$ from the three-dimensional (3D) power spectrum of the coarse-grained density field as $S(\bm{q},t)=\rho_{\bm q}(t) \rho_{-\bm q}(t)/N$. 
Here, the density field is defined as 
$\rho(\bm{r},t)= \sum_n \psi_n(a-|\bm r- \bm R_n(t)|)$, where $\psi_{n}({\bm r})$ is the hyperbolic tangent function and $\bm{R}_n $ is the position of the $n$-th particle. 
The characteristic wavenumber $\langle q \rangle$ corresponds to the first moment of $S(q,t)$ and is calculated as $\langle q \rangle = \int {dq} \ q S(q,t)/\int dq \ S(q,t) $.

\subsection*{Structure relaxation time~$\tau_\alpha$}

We calculate the structure relaxation time~$\tau_\alpha$, representing 
the characteristic time scale of particle rearrangement within the dense network phase.
We conduct independent simulations of a bulk phase where the volume fraction is
obtained by averaging around the peak of the distribution function of $\pi\sigma^3/6V_{\rm vor}$, where $V_{\rm vor}$ is the local Voronoi volumes of particles. 

We calculate $\tau_\alpha$ for a binary charged bulk PE solution with chain lengths $(N_\mathrm{c}, N_\mathrm{a}) = (40,40)$ at Bjerrum lengths $l_{\mathrm{B}} = 2\sigma$ (volume fraction $\phi \approx 0.38$) and $l_{\mathrm{B}} = 3\sigma$ (volume fraction $\phi \approx 0.42$), using FPD simulations (system size $L = 34.6\sigma$).
The initial configuration is prepared by sufficiently long BD simulations with 
full electrostatic interactions and WCA interactions (cutoff~$r_{\rm cut}=2^{1/6}\sigma$). 

We then include hydrodynamic interactions using the FPD method and further equilibrate the system for around $200\tau_{\rm BD}$. We compute the self-intermediate scattering function, $F_{s}({q}, t) = (1/N) \sum_{j=1}^{N} \exp(-\mathrm{i} {\bm q} \cdot ({\bm r}_{j}(t) - {\bm r}_{j}(0)))$, where the vectors ${\bm q}$ are uniformly distributed on a sphere of radius $|{\bm q}| = q$, and $q$ is the first peak of the structure factor $S(q)$.

The function $F_{s}({q}, t)$ is used to estimate the value of $\tau_\alpha$, which we define as the time at which $F_{s}({q}, t)$ decays to $e^{-1}$.

\subsection*{Characteristic domain deformation time~$\tau_\mathrm{d}$}

To estimate the characteristic domain deformation time, we compute the strain tensor field $\varepsilon_{\alpha \beta}$ from a coarse-grained displacement field, following the approach in our previous work~\cite{tateno2021}. This field is defined as ${\bm u}({\bm r}, t) = \left( \sum_{i} {\bm u}_{i}(t) G\left({\bm r} - {\bm R}_{i}\right) \right) / \left( \sum_{i} G\left({\bm r} - {\bm R}_{i}\right) \right)$, where $G({\bm r}) = \exp \left(-{{\bm r}^{2}}/{2\sigma^{2}}\right)$ is a Gaussian filter and ${\bm u}_{i}(t) = {\bm R}_{i}(t+\delta t) - {\bm R}_{i}(t)$ is the displacement of the $i$-th particle from time $t$ to time $t+\delta t$.

The strain tensor $\varepsilon_{\alpha \beta}$ is given by $\varepsilon_{\alpha \beta}({\bm r}, t) = \frac{1}{2} \left( \frac{\partial u_{\alpha}({\bm r}, t)}{\partial r_{\beta}} + \frac{\partial u_{\beta}({\bm r}, t)}{\partial r_{\alpha}} \right)$. We use the central differencing scheme to calculate the derivative, namely $\frac{\partial u_{\alpha}({\bm r}, t)}{\partial r_{\beta}} \approx \frac{u_{\alpha}({\bm r}+\Delta{\bm e}_\beta, t) - u_{\alpha}({\bm r}-\Delta{\bm e}_\beta, t)}{2\Delta}$, where $\Delta = 3l_0 \approx 0.5\sigma$ and ${\bm e}_\beta$ is the unit vector in the $\beta$-direction.

The sum of the diagonal components $\varepsilon_{\rm bulk} = \varepsilon_{xx} + \varepsilon_{yy} + \varepsilon_{zz}$ gives the bulk strain, while the sum of the non-diagonal components $\varepsilon_{\rm shear} = \varepsilon_{xy} + \varepsilon_{yz} + \varepsilon_{xz}$ refers to the shear strain. For short elapsed times $\delta t$, both $|\varepsilon_{\rm bulk}|$ and $|\varepsilon_{\rm shear}|$ show a linear increase with $\delta t$, from which we can estimate the strain rate as $|\varepsilon_{\rm bulk}|/\delta t$ and $|\varepsilon_{\rm shear}|/\delta t$. The characteristic times of domain deformation are then estimated as the inverse of the strain rates, $\delta t/|\varepsilon_{\rm bulk}|$ and $\delta t/|\varepsilon_{\rm shear}|$.

To reduce the influence of thermal noise, we average the configurations ${\bm R}{i}(t+\delta t)$ and ${\bm R}{i}(t)$ over a short time interval of $0.8\tau_{\rm BD}$ as in our previous work~\cite{yuan2023mechanical}.

\subsection*{Distance from counterion/monomer to network surface} 

To analyse the distribution of counterions and monomers around the network surface, we employ a Gaussian filter $G({\bm r})=\exp \left(-{{\bm r}^{2}}/{2\sigma^{2}}\right)$ to generate  a continuous density field defined as 
$\rho(\bm{r},t)= \sum_n G(\bm r- \bm R_n(t))$. We define the region of space where $\rho({\bm r}, t)\ge\rho_{\text {th }}$ (with $\rho_{\text {th}}$ set to 0.5) as the network phase. For a counterion situated in the void phase ($\rho({\bm r}, t)<\rho_{\text {th}}$), the distance~d to the surface is determined by the minimum distance between the counterion and any of the grid points where $\rho({\bm r}, t)\ge\rho_{\text {th}}$. Conversely, for a monomer within the dense network, the distance~d to the surface is determined by the minimum distance between the monomer and any grid point where $\rho({\bm r}, t)<\rho_{\text {th}}$.

While the choice of cutoff $\rho_{\text {th}}$ may slightly affect the data in Fig.~\ref{fig:Fig3} and Fig.~\ref{fig:Fig4}, we confirm that the underlying physics remains unchanged. We have also examined the results using a different cutoff value, $\rho_{\text {th}}=0.4$, as shown in Supplementary Fig.~S2. These results validate the robustness of our physical insights. Specifically, in the case of charge symmetry, we observe identical distributions for both cations and anions, as well as for the monomers of polycations and polyanions, supporting an overall charge-neutralised network. Conversely, in the case of charge asymmetry, a predominant net charge appears on the outer surface of the network, leading to a dynamic slowing down of domain coarsening.


%

\end{document}